\documentclass[a4paper,fleqn,usenatbib]{mnras}
\usepackage[T1]{fontenc}
\usepackage{ae,aecompl}
\usepackage{amsmath}
\usepackage{amsfonts}
\usepackage{amssymb}
\usepackage{graphicx}
\usepackage{color,xcolor}
\def\mnras{MNRAS}
\def\aap{A$\&$A}
\def\nat{Nature}
\def\apj{ApJ}
\def\apjl{ApJ Letters}
\def\apjs{ApJS}

\def\aj{AJ}

\newcommand{\me}{\, {\rm M}_{\oplus}}
\newcommand{\msun}{\, {\rm M}_{\odot}}

\newcommand{\au}{\, {\rm au}}

\title[Disc Mass versus UV ages]{How the gradient of $M_{\rm d}$ versus UV field strength yields insights into the ages of protoplanetary disc populations}
\author[G. A. L. Coleman \& S. van Terwisga]{Gavin A. L. Coleman$^{1}$\thanks{Email: gavin.coleman@qmul.ac.uk} \& Sierk E. van Terwisga$^{2}$\\
1. Astronomy Unit, Department of Physics and Astronomy, Queen Mary University of London, Mile End Road, London, E1 4NS, UK\\
2. Space Research Institute, Austrian Academy of Sciences, Schmiedlstrasse 6, A-8042 Graz, Austria}

\date{Accepted 2025 November 5. Received 2025 October 28; in original form 2025 May 29}
\pubyear{2025}
\begin{document}
\label{firstpage}
\pagerange{\pageref{firstpage}--\pageref{lastpage}}
\maketitle
\begin{abstract}
FUV radiation from massive stars launch photoevaporative winds from the outer regions of protoplanetary discs around other stars, removing gas and dust. Observations have identified a relation between the median dust disc mass and the external UV field strength. Here we use disc evolutionary models to explore how this relation evolves over time, and with respect to other stellar and disc properties. We find that the slope for the relationship $\lambda_{\rm UV}$ flattens over time as populations age, possibly explaining the differences seen between the L1641-N and L1641-S clusters in Orion A. We determine that $\lambda_{\rm UV}$ depends on the stellar mass where more massive stars exhibit steeper gradients than their lesser counterparts, in agreement with the differences seen between Herbig and T Tauri stars. Additionally, the strength of the mechanism for angular momentum transport, either viscosity or MHD disc winds, is found to significantly affect $\lambda_{\rm UV}$ with stronger $\alpha$ values reducing $\lambda_{\rm UV}$ due to more material accreting on to the central stars in weaker UV environments. Estimates of $\lambda_{\rm UV}$ from observations of L1641 place preliminary constraints on $\alpha$ to be between $10^{-3.5}$--$10^{-2.5}$, consistent with literature estimates. Further observations in different regions and better classifications of stellar masses will allow us to place stringent constraints on disc evolution properties, improving our understanding of how protoplanetary discs evolve.
\end{abstract}

\section{Introduction}
\label{sec:intro}

Planets and planetary systems are thought to arise from the evolution of circumstellar discs of gas and dust around young stars \citep[e.g.][]{Andrews18,Keppler18,Teague19}. These protoplanetary discs have been frequently observed in young star forming regions such as Orion \citep{PastPresentFuture2025}, where a number of trends have been uncovered.
These trends generally involve correlating observables such as disc masses, disc sizes, mass accretion rates, and age with each other \citep[see for example][for a review]{Manara23}.

Within these young star forming regions, massive stars emit large amounts of high energy radiation that launch photoevaporative winds from protoplanetary discs \citep{Odell94,Richling00,Adams04,Kim16,Haworth19}. These winds have been shown to play an important role in setting the evolutionary pathway of protoplanetary discs \citep{Coleman22}, their masses \citep{Mann14,Ansdell17}, radii \citep{Eisner18} and lifetimes \citep{Guarcello16,ConchaRamirez19,Sellek20}, even in weak UV environments \citep{Haworth17,Coleman25b}.
More recently, the role of high-energy radiation has been explored and determined that disc masses decrease as the proximity to massive stars increases \citep{Mann10,Ansdell17,VanTerwisga19,VanTerwisga20,Mauco23}. This trend highly depends on the radiation field around a massive star as theoretical work has shown that disc masses and disc lifetimes are larger for stars in low UV environments compared to those in high UV environments \citep{Johnstone98,Storzer99,Richling00,Scally01,ColemanHaworth20,Coleman22,Haworth23,Coleman24MHD,Coleman25b}. Indeed a trend to this effect was seen in the southern part of Orion A, where median dust disc masses were seen to decrease as the ambient UV field strength increased \citep{VanTerwisga23}. Whilst the mass of discs in $\sigma$ Orionis agreed with this trend \citep{Ansdell17}, those in the more extreme radiation environment the ONC bucked it, where their disc masses were higher than expected, \citep{Eisner18}. One possible explanation for this could be the ages of those clusters, and even the subgroups in them. Whether such trends are also dependent on age has yet to be explored, both in observations and in theoretical models.

To explain the trends and evolution of protoplanetary discs, 1D disc models have been developed including as many physical processes as possible. These processes determine how, and on what time-scales protoplanetary discs evolve. For example discs lose mass through accretion on to the central star \citep[e.g.][]{Pringle81,Manara20,Manara21}, internal photoevaporative \citep[e.g.][]{Alexander07, Owen12, Ercolano17} or magnetically driven winds \citep{BalbusHawley1991, Suzuki09, Pascucci23} as well as winds driven by external irradiation from nearby massive stars \citep[e.g.][]{Odell94, Richling00, Adams04, Kim16, Haworth19,Haworth21}. By including these processes, protoplanetary disc models have recently explored when and how the different processes dominate the evolution pathways of protoplanetary discs \citep[e.g.][]{ColemanHaworth20,Coleman22,Coleman24MHD}. Whilst the models are able to utilise those processes, they do however have to make assumptions on a number of the parameters. For example, the accretion prescriptions on to the central star, through either viscosity \citep{Shak} or MHD disc winds \citep{Tabone22} utilise an $\alpha$ parameter, which can be measured as a form of angular momentum transport within the discs. For viscous discs, this corresponds to the strength of the turbulence within protoplanetary discs \citep{Shak}, whilst it represents the strength of the MHD wind for wind driven discs \citep{Tabone22}. Observations of nearby star forming regions have been able to place constraints on values of $\alpha$ required for matching stellar accretion rates as well as the vertical extent of dust discs \citep{King07,Flaherty17,Ansdell18,Trapman20,Villenave20,Villenave22}. They typically find a range in $\alpha$ from $\sim10^{-4}$ to $\sim 10^{-2}$ with a median between $3\times10^{-4}$--$3\times10^{-3}$ \citep[see][for a recent review]{Rosotti23}.

In this work, we generate populations of discs to explore what drives the observed log-linear trends seen in \citet{VanTerwisga23} where dust disc masses decrease with increasing UV field strength. We first determine that our disc models are capable of reproducing the observed trends across a wide parameter space, before then showing how the gradient between disc mass and UV field strength evolves over time. Such an evolution of the gradient could then be used as an indicator of age, where for the L1641 cluster of YSOs, the North region would be predicted to be older than the South region due to the flatter gradients. Finally we show that the evolution of the gradients depends on the mass flow rate through disc, which allows observations of the gradients to also place further constraints on the viscosity $\alpha$ parameter or the strength of MHD winds.

\section{Population of simulated discs}
\label{sec:disc_model}

In order to compare with the observed features in protoplanetary discs it is necessary to generate populations of evolving discs. To do this we use the disc model outlined in \citet{Coleman24MHD} that evolves protoplanetary discs including accretion on to the central star through either viscosity \citep{Shak} or MHD disc winds \citep{Tabone22}, and mass lost from the disc through photoevaporative winds including both those internally driven due to high energy X-rays emanating from the central star \citep{Picogna19,Ercolano21,Picogna21}, and also those externally driven due to UV photons originating from massive stars in the local star forming environment \citep{Haworth23}. Including all of these effects, the gas surface density of each disc is evolved by
\begin{equation}
\label{eq:diffusion}
\begin{split}
    \dot{\Sigma}(r)=&\dfrac{1}{r}\dfrac{d}{dr}\left[3r^{1/2}\dfrac{d}{dr}\left(\nu\Sigma r^{1/2}\right)\right]+\dfrac{3}{2r}\dfrac{d}{dr}\left[\dfrac{\alpha_{\rm DW}\Sigma c_{\rm s}^2}{\Omega}\right]\\
    &-\dfrac{3\alpha_{\rm DW}\Sigma c_{\rm s}^2}{4(\lambda-1)r^2\Omega}-\dot{\Sigma}_{\rm PE}(r)
\end{split}
\end{equation}
where $\nu=\alpha_{\rm \nu} H^2\Omega$ is the disc viscosity with viscous parameter $\alpha_{\rm v}$ \citep{Shak}, $H$ is the disc scale height and $\Omega$ the Keplerian frequency.
The second and third components on the right hand side of eq. \ref{eq:diffusion} represent the change in surface density due to the angular momentum extracted by MHD wind, and the mass lost in the MHD wind itself.
The fourth term represents mass extracted by photoevaporative winds, which we discuss below.
Note there are two components for $\alpha$ in eq. \ref{eq:diffusion}, those being the measure of turbulence for viscosity $\alpha_{\rm \nu}$, and the measure of angular momentum extracted in the wind $\alpha_{\rm DW}$. In this work, we explore discs that are either viscous, or MHD wind driven, with one parameter taking the full value of $\alpha$, and the other being set to 0.
We evolve each disc for 20 Myr, or until the remaining gas mass is less than 0.1$\me$, at which point we assume the disc is fully dispersed.

To create a population of discs, we must first generate a population of stars. To do this we randomly draw stellar masses $M_*$ from a \citet{Kroupa01} initial mass function
\begin{equation}
    \xi (M_*)\propto \left\{ \begin{array}{ll}
M_*^{-1.3} & {\rm for}~ 0.08\msun\le M_* < 0.5\msun \\
\\
M_*^{-2.3} & {\rm for}~ 0.5\msun\le M_* < 10\msun.
\end{array} \right.
\end{equation}
For each star we then determine it's X-ray luminosity by following the linear relation found in \citet{Flaischlen21}, by randomly drawing a value from a normal distribution where the mean is equal to 
\begin{equation}
    \log{\overline{L_X}} = 30.58 + 2.08\log{\left(\frac{M_*}{\msun}\right)} 
\end{equation}
with a standard deviation of 0.29. For each population we vary the $\alpha$ value for either viscosity or MHD disc winds, as well as the external UV field strength. The initial disc mass is randomly chosen to be between 0.2--1$\times M_{\rm d, max}$ where $M_{\rm d,max}$ is the maximum disc mass that a gas disc of radius $r_{\rm ini}$ around a star of mass $M_*$ can be before becoming gravitationally unstable \citep{Haworth20}, and is equal to
\begin{equation}
    \label{eq:max_disc_mass}
    \dfrac{M_{\rm d, max}}{M_*} < 0.17 \left(\dfrac{r_{\rm ini}}{100\au}\right)^{1/2}\left(\dfrac{M_*}{\msun}\right)^{-1/2}.
\end{equation}
We initialize our discs following \citet{Lynden-BellPringle1974}
\begin{equation}
    \Sigma(r) = \Sigma_0\left(\frac{r}{1\au}\right)^{-1}\exp{\left(-\frac{r}{r_{\rm c}}\right)}
\end{equation}
where $r_{\rm c}$ is the scale radius that we take to being $r_{\rm c}=50\au \sqrt{M_*}$.
For each population we generate 1000 discs, with a specific value of $\alpha$ and UV field strength. We vary $\alpha$ between $10^{-4}$--$10^{-2}$, and the external UV field between $10^{1.25}$ and $10^{3.5} \rm G_0$.

\begin{figure}
\centering
\includegraphics[scale=0.55]{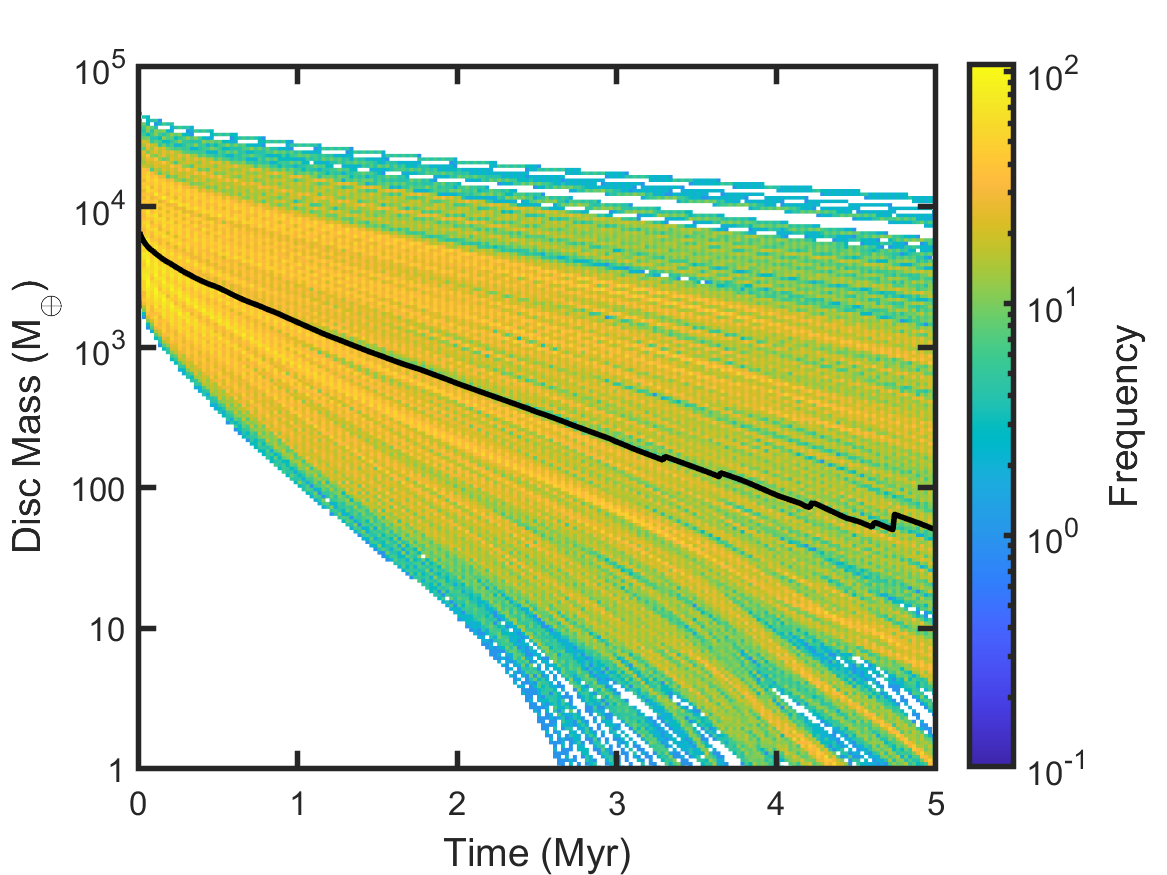}
\caption{Disc mass evolution tracks for a single population of viscously evolving ($\alpha=10^{-3}$) protoplanetary discs in a $10^{3} \rm G_0$ environment. The colour shows the frequency of discs with those masses as they temporally evolve. The black line shows the median disc mass.}
\label{fig:discmass_evolution}
\end{figure}

\begin{figure*}
\centering
\includegraphics[scale=0.55]{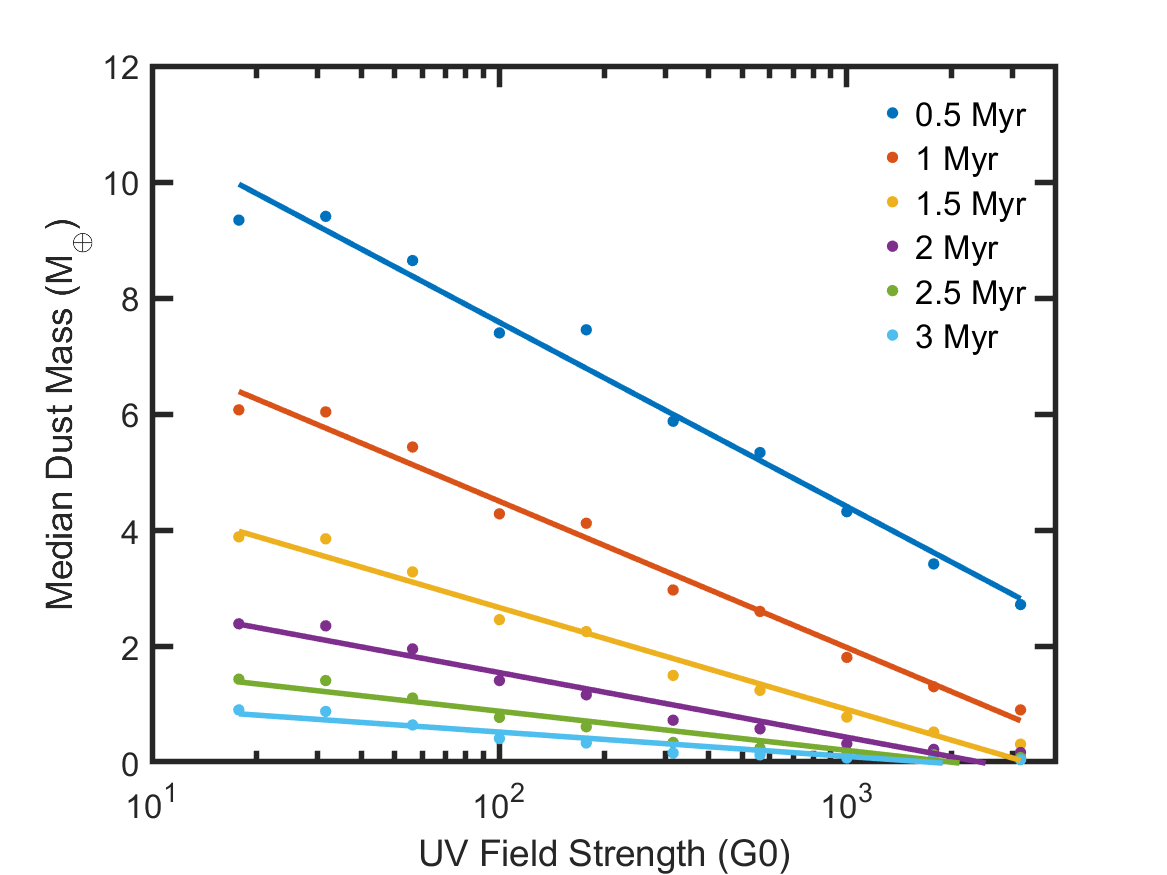}
\includegraphics[scale=0.55]{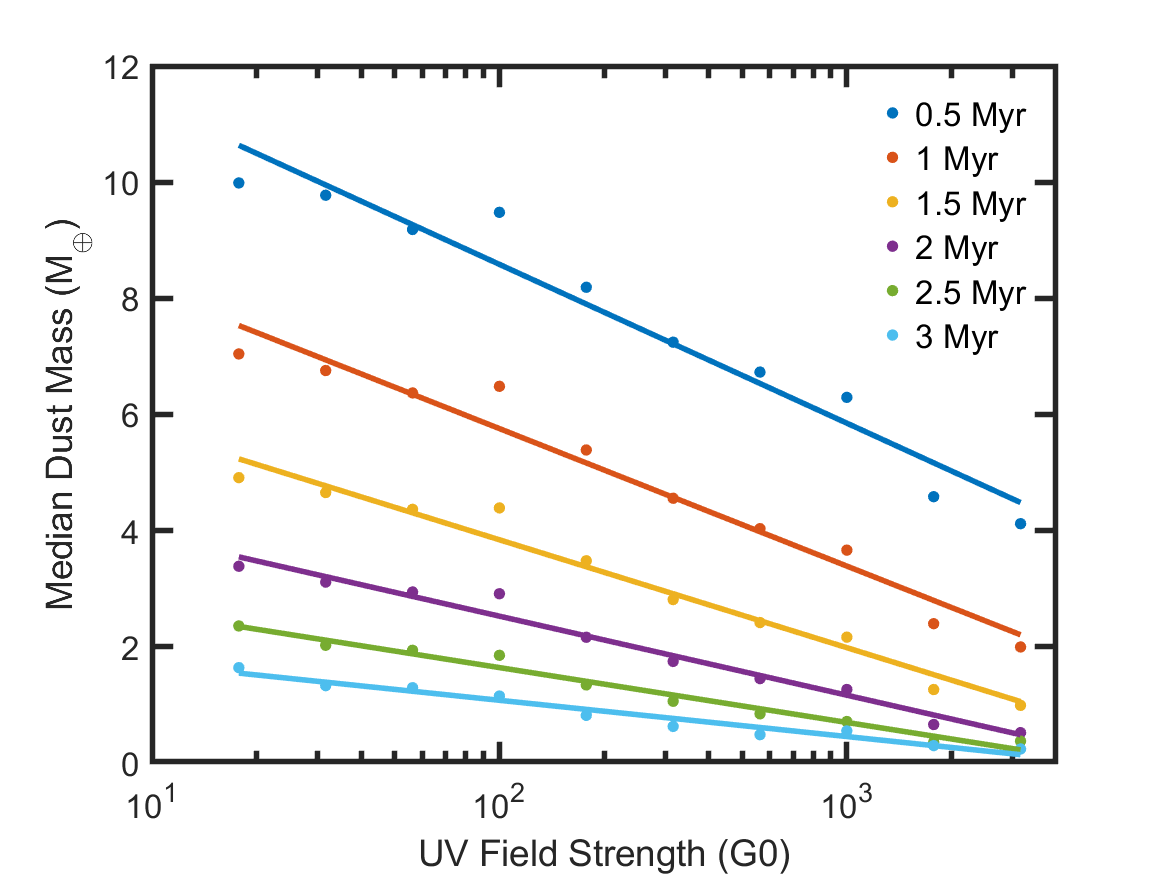}
\caption{Median disc mass for a population of stars as a function of the UV field strength. Different colours represent different ages for the populations ranging between 0.5--3 Myr. The left-hand panel shows the relations for evolving viscous discs, whilst the right-hand panel represents MHD wind driven discs. The lines represent the fits to the data following equation \ref{eq:fit}.}
\label{fig:pop_trends}
\end{figure*}

\section{Evolution of dust disc masses across different UV regimes}
\label{sec:results}

As discs evolve they lose mass through accretion on to the central star and via photoevaporative winds. In Fig. \ref{fig:discmass_evolution} we show the evolution of the disc mass over time for a population of discs evolving in a $10^3 \rm G_0$ environment. The discs evolved through viscous accretion with $\alpha=10^{-3}$. The colour coding shows the frequency of discs with those masses over time. The general trend in disc mass is clear here, with all discs reducing in mass over time. Some discs lose mass more quickly than others, such as those discs that are initially less massive, but also around stars with stronger X-ray luminosities. Additionally, the spread in stellar masses also contributes to the spread in disc masses and their subsequent evolution since lower mass stars are relatively more strongly affected by external photoevaporation of equal levels compared to more massive stars \citep{Coleman22,Haworth23}. The black line in Fig. \ref{fig:discmass_evolution} shows the median disc mass as a function of time, where the consistent decrease over 5 Myr is clear.

The evolution of disc masses in Fig. \ref{fig:discmass_evolution} only represented a single population in a single UV environment. However, star forming regions contain a range of UV fields depending on where stars form in respect to the resident massive stars. As shown in \citet{VanTerwisga23}, two regions in Orion, L1641-N and L1641-S exhibit ranges in UV fields spanning 1-2 orders of magnitude.
Additionally they found a relation between the strength of the UV field and the median disc mass, with the disc mass decreasing with the UV strength,
\begin{equation}
\label{eq:fit}
M_{\rm dust,median}= \lambda_{\rm UV}\log_{10} (F_{\rm FUV}/ {\rm G_0}) + C_{\rm UV}.
\end{equation}
with $\lambda_{\rm UV}$ = $-1.2_{-0.17}^{+0.16}$, and $C_{\rm UV}$ = $3.4_{-0.20}^{+0.19}$.
This was inferred from the properties of the discs in L1641 with UV fields between 1--1000 $\rm G_0$. Interestingly, whilst the subsamples L1641-N and L1641-S broadly agreed with the relation, the gradient $\lambda_{\rm UV}$ differed slightly, possibly indicating added complexity within the empirical relation across the region.

To compare with the trends found in \citet{VanTerwisga23}, we generate multiple populations of discs varying the background UV field strength for each. We explore UV field strengths between $10^{1.25} \rm G_0$--$10^{3.5}\rm G_0$. In Fig. \ref{fig:pop_trends} we show the median dust mass for the populations as a function of the UV field strength. The different colours represent the distributions at different disc ages. The left panel shows the trends for discs including viscous accretion, whilst the right hand panel shows them for MHD wind driven discs. Understandably, as time progresses, the median dust mass in both panels decreases, as dust is lost to the central star through radial drift, and through being entrained in photoevaporative winds \citep{Hutchison21}. Additionally as the background UV field increases, the dust mass also decreases as the discs are more rapidly truncated due to stronger external photoevaporative winds \citep{Coleman22}.

When comparing the accretion regime in the discs, i.e. viscosity to MHD disc winds, it is clear in Fig. \ref{fig:pop_trends} that qualitatively there is little difference between the two regimes, consistent with expectations from previous studies \citep{Coleman24MHD}. The dust disc masses for both regimes decrease as a function of the UV field with similar gradients. It is not unexpected that there would be little difference between the two regimes, as with the majority of the mass being in the outer disc regions where evolution through external photoevaporation dominates \citep{Coleman24MHD}, the role of how material accretes through the disc is not significant. The main effect that the accretion regimes have in the outer disc is on determining where and how quickly the disc truncates down to specific radii due to the magnitude of viscous expansion of the discs, but as was shown in previous works, their effects are negligible when internal and external photoevaporation processes are included \citep{Coleman24MHD}.

However, whilst qualitatively there is little difference, the rate of the decrease in disc mass over time is subtly different, with MHD wind driven discs evolving slower than their viscous counterparts, typically for this $\alpha$ value on order of 0.5 Myr. This is a result of weaker accretion on to the central stars for MHD wind driven discs with the same value of $\alpha$ and so more dust is retained in the discs for longer. Naturally, increasing the value of $\alpha$ would amend this, with the equivalent $\alpha$ value yielding qualitatively and quantitatively similar results to viscous disc models.

\begin{figure*}
\centering
\includegraphics[scale=0.55]{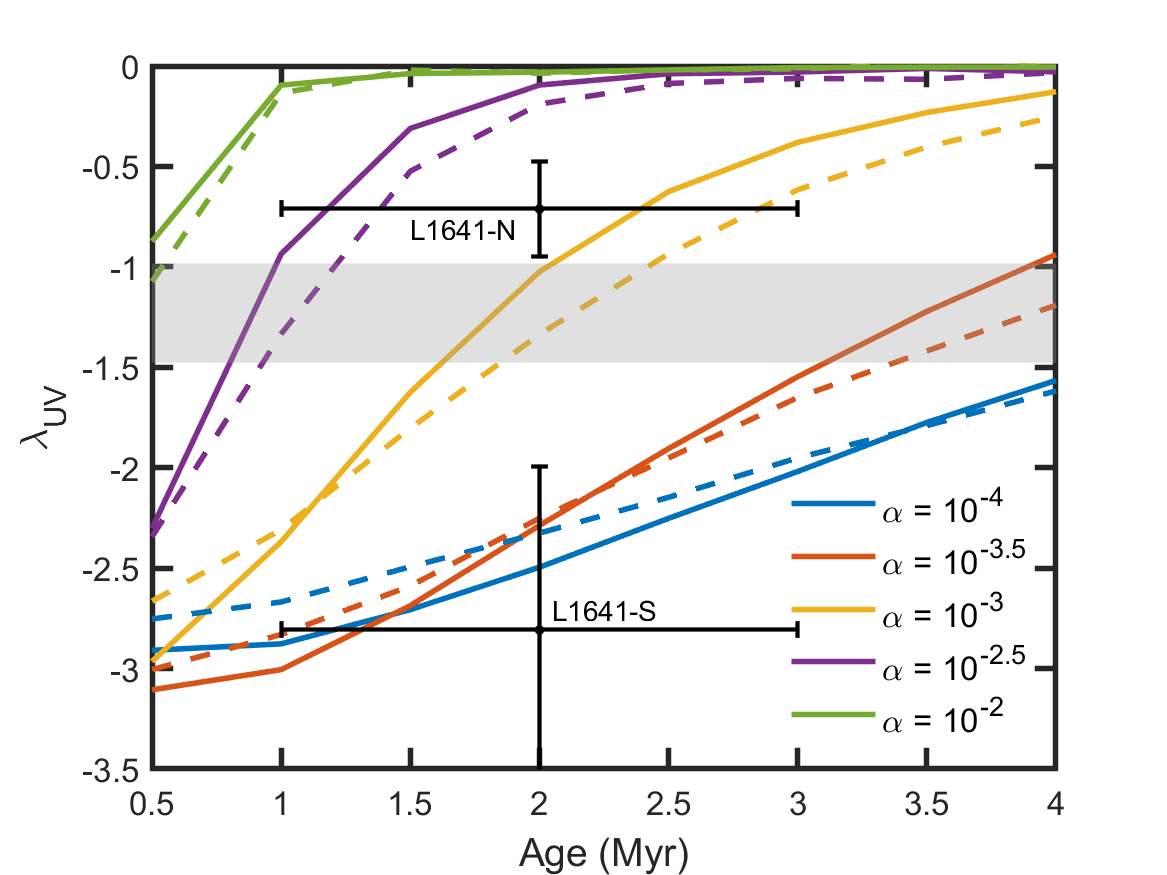}
\includegraphics[scale=0.55]{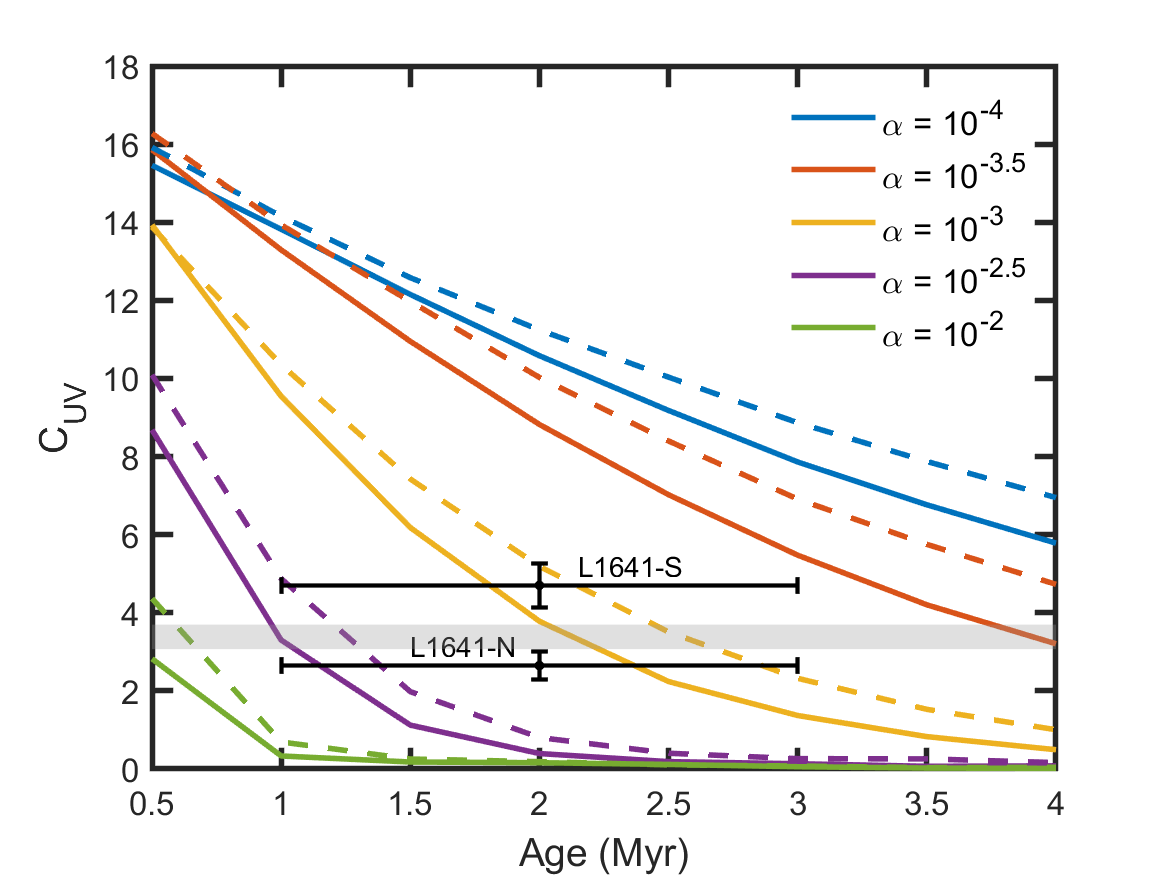}
\caption{The gradient $\lambda_{\rm UV}$ (left-hand panel) and the intercept $C_{\rm UV}$ (right-hand panel) of the relation between the strength of the UV field and the median disc mass. Different colours denote different values of $\alpha$ for the simulated discs, with $\alpha=10^{-4}$ (blue), $\alpha=10^{-3.5}$ (red), $\alpha=10^{-3}$ yellow), $\alpha=10^{-2.5}$ (purple) and  $\alpha=10^{-2}$ (green). Solid lines represent viscous discs, whilst dashed lines represent MHD wind driven discs. The grey patch represents the range in $\lambda_{\rm UV}$ and $C_{\rm UV}$ found by \citet{VanTerwisga23}, with the black points showing the values found for specific regions.}
\label{fig:gradients}
\end{figure*}

\subsection{Role of age in determining $\lambda_{\rm UV}$ and $C_{\rm UV}$}
\label{sec:age}

A common feature for both viscous and MHD wind driven discs, is that $\lambda_{\rm UV}$ and $C_{\rm UV}$ vary with age. Figure \ref{fig:gradients} shows the temporal evolution of $\lambda_{\rm UV}$ (left-hand panel) and $C_{\rm UV}$ (right-hand panel) for viscous (solid lines) and MHD wind driven discs (dashed lines). The colours show different values of $\alpha$. At the start of the disc lifetimes, both $\lambda_{\rm UV}$ and $C_{\rm UV}$ are large, due to the initial discs being larger around more massive stars. As they evolve, the discs around more massive stars lose more mass to photoevaporation and accretion than those around low mass stars. This acts to flatten out the gradient of disc mass versus UV field strength, and is clearly seen by the yellow line showing the discs that evolved with $\alpha=10^{-3}$ (similar to those described above), where the gradient flattens as time progresses.
The intercept $C_{\rm UV}$ also decreases as the discs age, since more mass is lost to photoevaporation and accretion on to the central stars.
Interestingly, the extent to which the gradient flattens and the intercept reduces over time varies with $\alpha$. For larger $\alpha$ values, the slope flattens and the intercept drops quickly as the gas discs are more rapidly depleted across all UV environments through accretion on to the central stars. This naturally reduces the magnitude of the gradients and intercepts as all of the discs are of lower mass. The opposite occurs for discs evolving with lower values of $\alpha$.

The grey patch in Fig. \ref{fig:gradients} shows the fit for the L1641 cluster \citep{VanTerwisga23}, showing good agreement across a range of $\alpha$ values depending on the cluster age. With the cluster age expected to be between 1--3 Myr, this translates to the observed gradients being consistent with $\alpha$ values between $10^{-3.5}$--$10^{-2.5}$, irrespective of whether the accretion through the discs is driven by viscosity or MHD winds. These values are consistent with other estimates for the underlying $\alpha$ parameter based on accretion rates on to central stars \citep{Ansdell18,Trapman20}, the radial extent of structures such as rings \citep{Dullemond18,Sierra19}, the radial extent of dust discs \citep{Toci21}, and the combination of matching accretion rates, external mass loss rates, disc sizes and masses \citep{Coleman25d}.

Interestingly, when splitting the L1641 cluster into two regions, those in the south and those in the north as was performed in \citet{VanTerwisga23}, we find that the two regions have distinctly different gradients and slightly different intercepts. This is clearly seen by the south region exhibiting a much steeper gradient in the median dust disc mass as a function of the UV field strength than the north, whilst having comparatively similar intercept values. This can be seen by the black points and their appropriate error bars in Fig. \ref{fig:gradients}. Should future observations confirm this trend, then that could imply two possible outcomes. Either the two regions have distinctly different levels of turbulence, or they are of different ages. Indeed it is thought that L1641-N may be older due to a foreground population of stars \citep{VanTerwisga22}, potentially linked to the location of the older NGC 1980 cluster \citep{Alves12,Zari19}. Future observations of these regions and clarification on their ages and gradients in disc mass will therefore aid in determining the strength of $\alpha$ required for simulations to match the observations.

\begin{figure}
\centering
\includegraphics[scale=0.55]{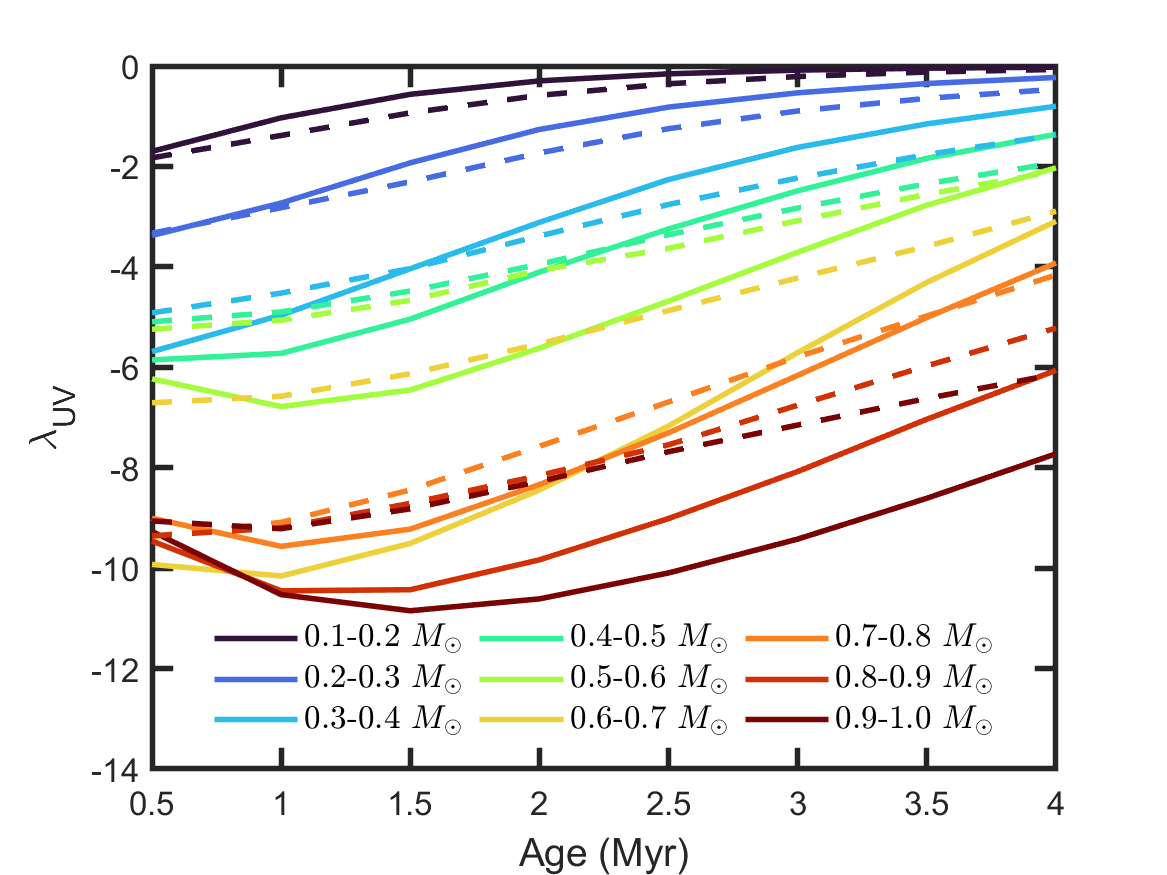}
\caption{Same as Fig. \ref{fig:gradients} but showing the gradients for populations of different stellar masses ranging from 0.1--0.2 $\msun$ to 0.9--1 $\msun$. The value of $\alpha$ was set to $10^{-3}$.}
\label{fig:gradients_star}
\end{figure}

\subsection{Changing gradient across stellar mass}

The populations above included stars across a range of masses between 0.1--1$\msun$, and showed that the age of the population affects the gradient in median disc mass as a function of the UV field strength. However if it is possible to observationally divide clusters into stellar mass bins, then there may be different gradients across those bins since more massive stars tend to have more massive discs than low mass stars. To test this, we computed an additional batch of populations, where each population contained 1000 stars with only the stellar mass of the stars within the populations changed. Each population covered a 0.1$\msun$ bin, ranging from between 0.1--1$\msun$. We then calculated the gradients of the median dust disc masses as a function of the UV field strength, to determine the impact of stellar mass. We again used viscous and MHD wind driven discs, but only focussed on $\alpha$ values of $10^{-3}$.

Figure \ref{fig:gradients_star} shows the values of $\lambda_{\rm UV}$ that arise for each population of stars, as a function of age. Similar to Fig. \ref{fig:gradients}, $\lambda_{\rm UV}$ starts off as a steep function for young stars, and flattens as the populations age. However, what is seen in Fig. \ref{fig:gradients_star} is that more massive stars contain much steeper gradients than their less massive counterparts. For example, $\lambda_{\rm UV}\leq-2$ for stars with masses less than 0.2$\msun$, whilst for stars with masses $0.9\le M_*\le 1\msun$, $\lambda_{\rm UV}\sim -9$ at 0.5 Myr, before then steepening to $\sim-11$ at 1.5 Myr as the discs in the more extreme UV environments are rapidly depleted, before then flattening out over the next 2.5 Myr to only be $\sim-8$ after 4 Myr. This indicates that the differences in $\lambda_{\rm UV}$ across different clusters may also be due to different stellar properties of the stars in those clusters. Indeed \citet{Mauco23} observe steeper gradients for stars with masses $M_* > 0.4\msun$ than for those $M_*<0.4\msun$ for a small selection of stars in $\sigma$ Orionis. Additionally when looking at Herbig stars in Orion, \citet{Stapper24} find that the relationship between the median dust disc mass and the UV field strength is $\lambda_{\rm UV}=-7.6$, much steeper than that found for T Tauri stars of $-1.3$ presented in \citet{VanTerwisga23}. This steepening as a function of stellar mass is in agreement with that seen in Fig. \ref{fig:gradients_star}.

Additionally, when calculating the full populations as was done in Sect. \ref{sec:age}, but by varying the lower limits for the stellar population, the trends closely follow that of the lowest mass stars, since they are the most populous. For example by using a lower stellar mass limit of 0.3$\msun$ instead of 0.1$\msun$, $\lambda_{\rm UV}$ moves from ranging between -3 to -0.2 for stars where $\alpha=10^{-3}$ to between -5 to -2, consistent with that seen in Fig. \ref{fig:gradients_star} for stars between 0.3--0.4$\msun$. This highlights the importance of understanding the mass distributions of stellar populations when attempting to determine the underlying properties affecting their evolution. Further observations of large numbers of stars with varied and precise stellar masses should therefore allow good constraints to be placed on the values of $\alpha$, and when combined with other estimates, could indicate whether protoplanetary discs evolve viscously or through MHD winds. Additionally, for the more massive stars, there is a more notable difference in the gradients between viscous and MHD wind driven discs, with viscous discs containing steeper gradients.

\section{Discussion and Conclusions}
\label{sec:conc}

Recent observations of populations of protoplanetary discs has unveiled a relation between the median dust disc mass and the strength of the external UV field strength \citep{VanTerwisga23}. In this work, we use protoplanetary disc evolution models to determine first whether they can match the observed relations, and then how the relation changes over time and with other protoplanetary disc properties. We draw the following main conclusions from this work:

(1) The relation between the median dust disc mass and the external UV field strength varies over time as protoplanetary discs evolve at different rates depending on the external UV field strength, and the initial stellar and disc properties.

(2) The gradient $\lambda_{\rm UV}$ varies over time as populations age, starting at steeper values at early ages, before flattening towards the end of all of the discs lifetime. The intercept $C_{\rm UV}$ also starts at higher values before consistently falling to smaller values.

(3) The level of turbulence or MHD wind strength also affects $\lambda_{\rm UV}$ and $C_{\rm UV}$ with stronger values of $\alpha$ yielding flatter slopes, since more material is accreted on to the central stars.

(4) The observed values for L1641 matches a wide range of values for $\alpha$ since the stellar ages are not well constrained. They predict $10^{-3.5}<\alpha<10^{-2.5}$ for the cluster as a whole. When splitting the cluster into North and South regions, the expected $\alpha$ values vary significantly, indicating that the two regions may be of different ages.

(5) Finally, the value for $\lambda_{\rm UV}$ depends on the stellar mass, with more massive stars exhibiting steeper slopes than low-mass stars, especially for viscous discs, consistent with observations comparing Herbig stars to T Tauri stars \citep{Stapper24}. This also affects the values for $\lambda_{\rm UV}$, within stellar populations as the lowest mass stars contribute the most in determining the value of $\lambda_{\rm UV}$. Therefore, understanding the mass distributions in stellar populations is necessary when attempting to determine the underlying properties affecting their evolution.

The models presented in this work ultimately show how the relation between median dust disc mass and the UV field strength depend on a number of properties, including stellar mass, age, and the level of turbulence or the MHD wind strength. They also show that with sufficient information on the age and stellar masses, then it is possible to place stringent constraints on internal protoplanetary disc processes such as viscosity and MHD wind strengths. Whilst for populations of stars it is difficult to differentiate between the two mechanisms as to which drives the exchange of angular momentum in protoplanetary discs, should precise constraints on stellar mass be obtained, then the relations found for higher mass stars could yield insights into whether discs are driven by viscosity or MHD winds. Additionally, by placing constraints on $\alpha$ for entire populations, will enhance the ability for other estimates of $\alpha$ to determine which process is dominant. This can only be achieved, if accurate data is provided on the stellar masses, UV field strengths, and the ages of stellar populations, in order to place tight constraints on $\lambda_{\rm UV}$ and $C_{\rm UV}$. Should that be accomplished, then those constraints would significantly aid in our understanding on protoplanetary disc evolution, and subsequently, how planets form.

\section*{Data Availability}
The data underlying this article will be shared on reasonable request to the corresponding author.

\section*{Acknowledgements}
The authors thank the anonymous referee for their useful and insightful comments that improved the quality of the paper.
GALC acknowledges funding from the UKRI/STFC grant ST/X000931/1.
This research utilised Queen Mary's Apocrita HPC facility, supported by QMUL Research-IT (http://doi.org/10.5281/zenodo.438045).
This paper makes use of the following ALMA data: ADS/JAO.ALMA\#2019.1.01813.S. ALMA is a partnership of ESO (representing its member states), NSF (USA) and NINS (Japan), together with NRC (Canada), MOST and ASIAA (Taiwan), and KASI (Republic of Korea), in cooperation with the Republic of Chile. The Joint ALMA Observatory is operated by ESO, AUI/NRAO and NAOJ.

\bibliographystyle{mnras}
\bibliography{references}{}

\label{lastpage}
\end{document}